\documentclass[11pt]{article}
\usepackage{moriond,epsfig}
\def\met{\mbox{${\hbox{$E$\kern-0.4em\lower-.1ex\hbox{/}}}_T$}} 
\newcommand{\Zee}{\mbox{$\rm Z\rightarrow\rm e^{+} \rm e^{-}$}}
\newcommand{\gammaZee}{\mbox{$\rm \gamma^{*}/Z\rightarrow\rm e^{+} \rm e^{-}$}}
\newcommand{\Wenu}{\mbox{$\rm W\rightarrow\rm e\nu$}}

\newenvironment{2figures}[1]{\begin{figure}[#1] 
  \begin{center}
    \begin{tabular}{p{.47\textwidth}p{.47\textwidth}} }
 {  \end{tabular}
  \end{center} 
 \end{figure}
}
\def\MyPhoto{\resizebox{3.5cm}{!}{\includegraphics{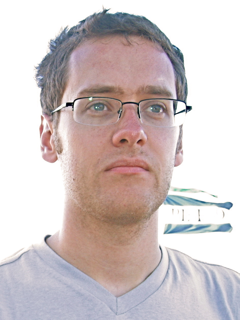}}}

\begin{document}
\vspace*{4cm}
\title{PREPARATIONS FOR MEASUREMENT OF THE W AND Z PRODUCTION CROSS-SECTIONS WITH EARLY CMS DATA}

\author{ D. R. WARDROPE }

\address{Department of Physics, Blackett Laboratory, Imperial College London,\\ Prince Consort Road, London, SW7 2BW, UK}

\maketitle\abstracts{The CMS analysis strategy for the early data measurement of the inclusive W and Z production cross-sections using their electron decay modes is presented. This measurement is expected to be among the first from the LHC and so appropriately robust selections and data-driven methods are used. A significant measurement is possible with an integrated luminosity of 10\,pb$^{-1}$.\\
\begin{center} \normalsize{\emph{Presented at XLIVth Rencontres de Moriond.}}\end{center}}
\section{Introduction}
The Compact Muon Solenoid\,\cite{CMSPaper} was designed to make discoveries at the TeV scale :  to elucidate the nature of electroweak symmetry breaking and to search for physics beyond the Standard Model. For any such discovery to be credible, it must first be demonstrated that the CMS detector and how it reconstructs events is understood; that the LHC environment is understood; and that backgrounds can be characterised and controlled. One mechanism to make these demonstrations is to measure well understood `standard candles', such as W and Z production. 

The measurement of the inclusive W and Z production cross-sections using the electron decay mode\,\cite{elecPAS} is particularly suitable for the early data. W and Z bosons are predicted to have large production cross-sections at the LHC, approximately 190\,nb and 60\,nb respectively\,\cite{SMatLHC}. Their experimental signature of well-isolated leptons with high transverse momenta ($p_T$)  is very distinctive in hadron collisions and should be readily triggered and selected. Thus, an integrated luminosity of only 10\,pb$^{-1}$ is sufficient for significant analyses of W and Z production.

A cross-section measurement made with 10\,pb$^{-1}$ will be one of the first results from CMS and the LHC. For this early data, the ultimate calibration and alignment of the detector will not be available and Monte Carlo simulations may not yet accurately describe the detector and the LHC environment. Thus the strategy for  making the measurements place emphasis on mitigating any effects consequent to this :  simple and robust selections are employed and data-driven methods are used to measure efficiencies and estimate signal and background yields.

\section{Reconstruction and Selection of W and Z Bosons}
The design of the CMS detector is based around a 4\,T large radius solenoid, containing the silicon-based inner tracking; the homogeneous, fully active, crystal electromagnetic calorimeter (ECAL); and the sampling hadronic calorimeter. Outside the solenoid are four layers of muon detectors, installed in the solenoid return yoke.

$W\rightarrow e\nu$ and $\gamma^*/Z\rightarrow e e$ events must pass the single isolated electron High Level Trigger requirements\,\cite{DAQTDR}. Further offline selection of $W\rightarrow e\nu$ requires one offline reconstructed electron within these events and $\gamma^*/Z\rightarrow e e$ requires two. An offline reconstructed electron\,\cite{PTDR1} consists of a supercluster in the ECAL, matched to a track from the interaction vertex. The supercluster is a collection of clusters of crystals, extended in the azimuthal direction to gather the energy radiated by an electron traversing the tracker. 

The reconstructed electrons in both event types must satisfy some robust identification criteria based on cluster shape and track-supercluster matching, which are designed to be efficient and effective at start-up. In order to select electrons characteristic of W and Z decay, the electrons must have high $p_T$ superclusters, with low activity around the electron is demanded in both the tracker and the calorimeters.

In $W\rightarrow e\nu$ events, the presence of the neutrino is inferred by an imbalance in the transverse energy vector sum of the event, \met. This missing transverse energy, calculated from calorimeter energy deposits, is used as a further discriminating variable for the estimation of signal and background event yields.
 \begin{2figures}{hbt}
   \resizebox{\linewidth}{!}{\includegraphics{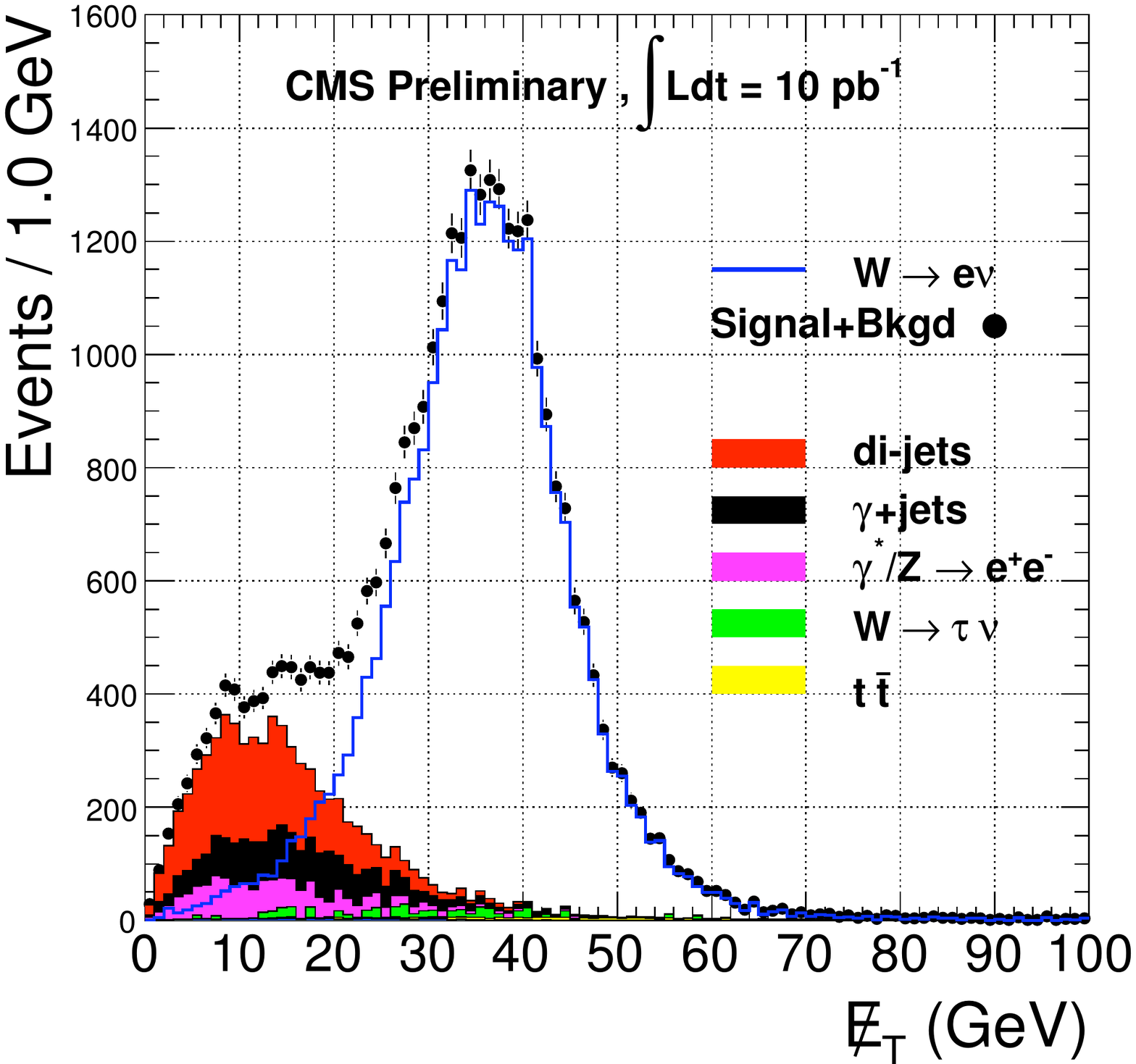}} &
   \resizebox{\linewidth}{!}{\includegraphics{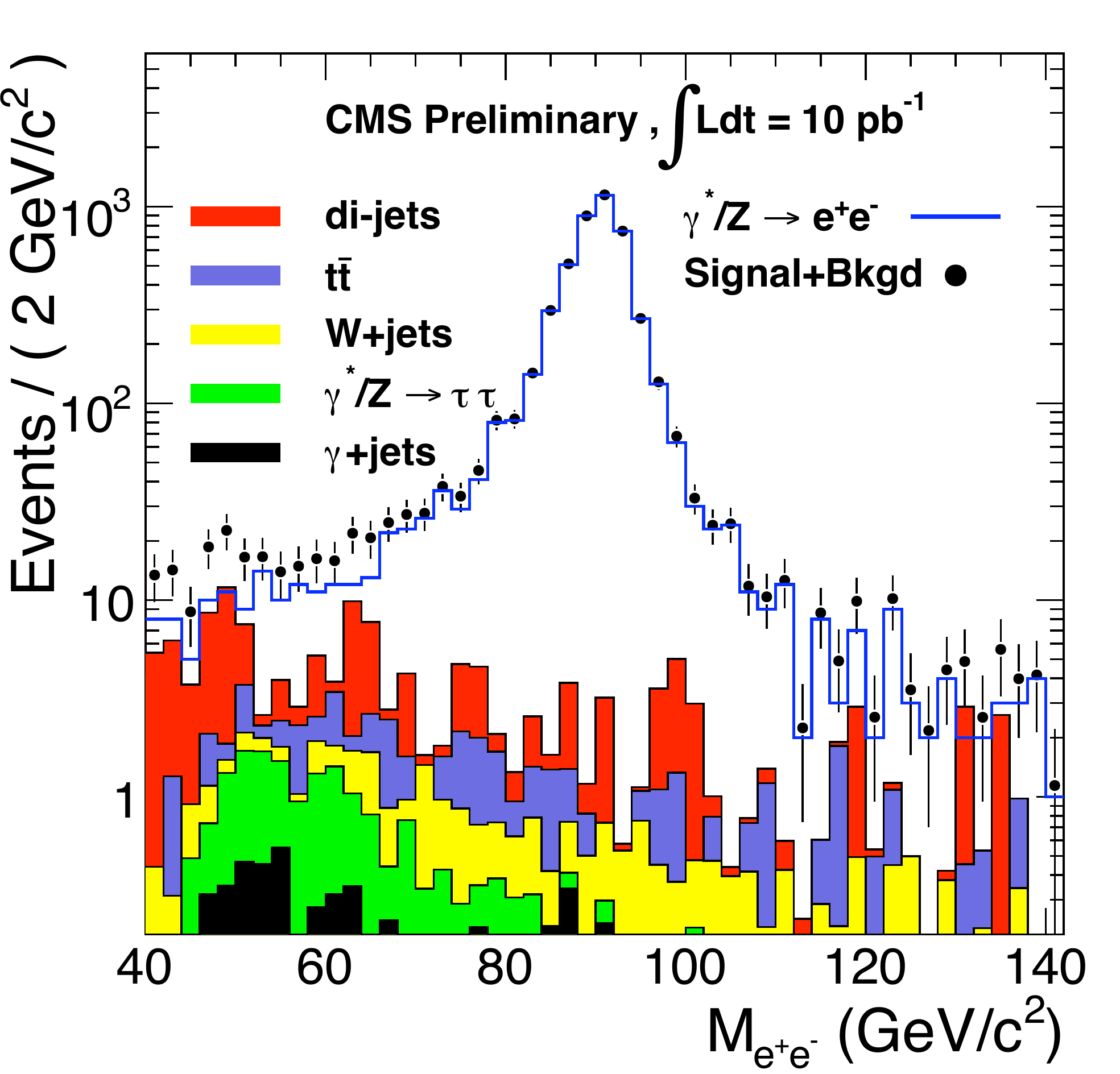}} \\
   \caption{\met distribution of  $W\rightarrow e\nu$ and its backgrounds, after selection, for 10\,pb$^{-1}$ of integrated luminosity. The largest background contributions are from QCD di-jet events.}
   \label{f_MEtDist_Elec} &
   \caption{Invariant mass distribution of  $Z\rightarrow ee$ and its backgrounds, after selection for 10\,pb$^{-1}$ of integrated luminosity. The requirement of two identified electrons heavily suppresses the backgrounds.}   \label{f_InvMassDist_Elec} \\
 \end{2figures}
 
\section{Efficiency Determination from Data}
The efficiency to reconstruct objects and to trigger and select events can be measured using the data-driven ``Tag and Probe'' method\,\cite{effPAS}. An unbiased and pure sample of leptons is obtained from \Zee for measuring the efficiency of a particular selection or reconstruction step. One electron, the `tag', meets stringent identification criteria to ensure it is an electron. The other, `probe', electron need satisfy only loose criteria and so is left unbiased. The purity of the probe sample is ensured by restricting the invariant mass of the electron pair to be about the Z mass.

\begin{figure}[hbt]
   \begin{center}
     \resizebox{6cm}{!}{\includegraphics{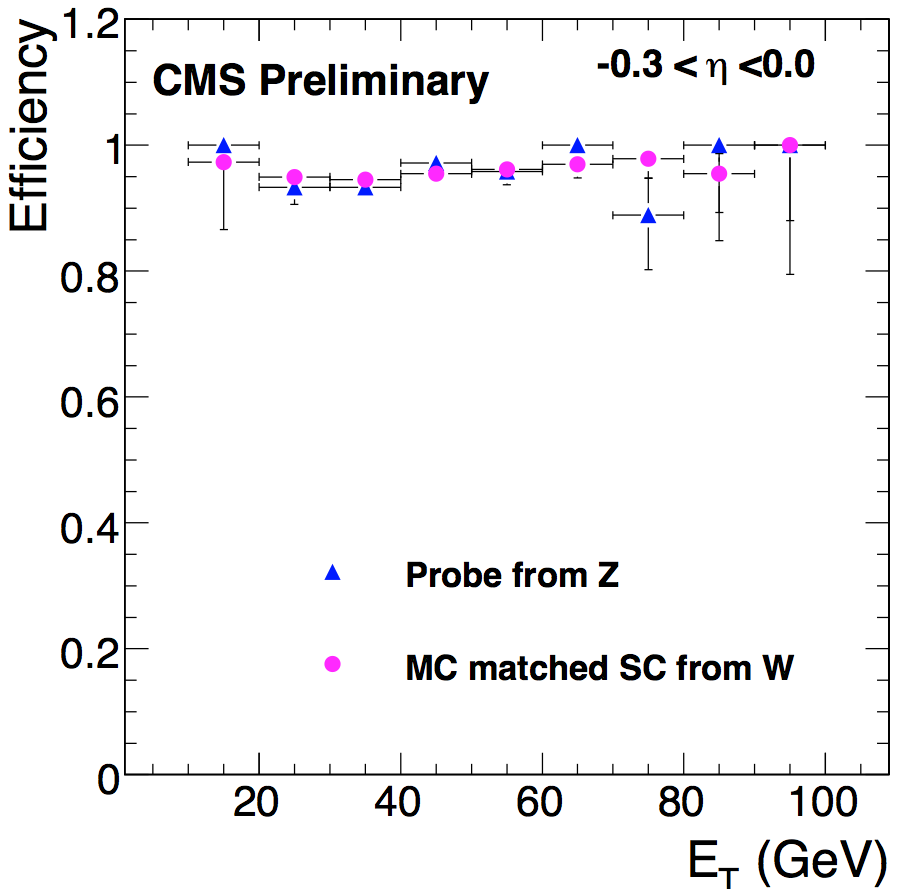}}
     \caption{Efficiency for an electron to have a track reconstructed and matched to its supercluster, as a function of the supercluster transverse energy. The efficiency determined with Tag and Probe (triangles) is compared to the true efficiency in \Wenu\ events (circles).}
     \label{f_TrackEff}
   \end{center}
 \end{figure}

The efficiencies measured using the Tag and Probe method have been validated against the true efficiencies from Monte Carlo simulations (Figure \ref{f_TrackEff}).

\section{Background Estimation}
Electroweak backgrounds in the W and Z samples are small and sufficiently well understood theoretically, so can be reliably estimated from simulation. However, the QCD di-jet background in \Wenu\ is much larger and intrinsically difficult to simulate. As a result background subtraction methods which do not rely on simulation will be used.

These methods require accurate descriptions, derived from data, of the properties of both signal and background events. For the background, such a description can be obtained by inverting one of the electron identification criteria applied in the analysis (Fig. \ref{f_dijetTemplate}). This has the effect of rejecting the signal and also the electroweak backgrounds, leaving a pure di-jet sample. For the signal, \Zee\ events (background free after selection) are made to represent \Wenu\ by removing the energy deposits of one electron from the \met\ sum. After corrections for neutrino acceptance and different boson masses, a good representation is obtained (Fig. \ref{f_Ersatz}).

\begin{2figures}{hbt}
   \resizebox{\linewidth}{!}{\includegraphics{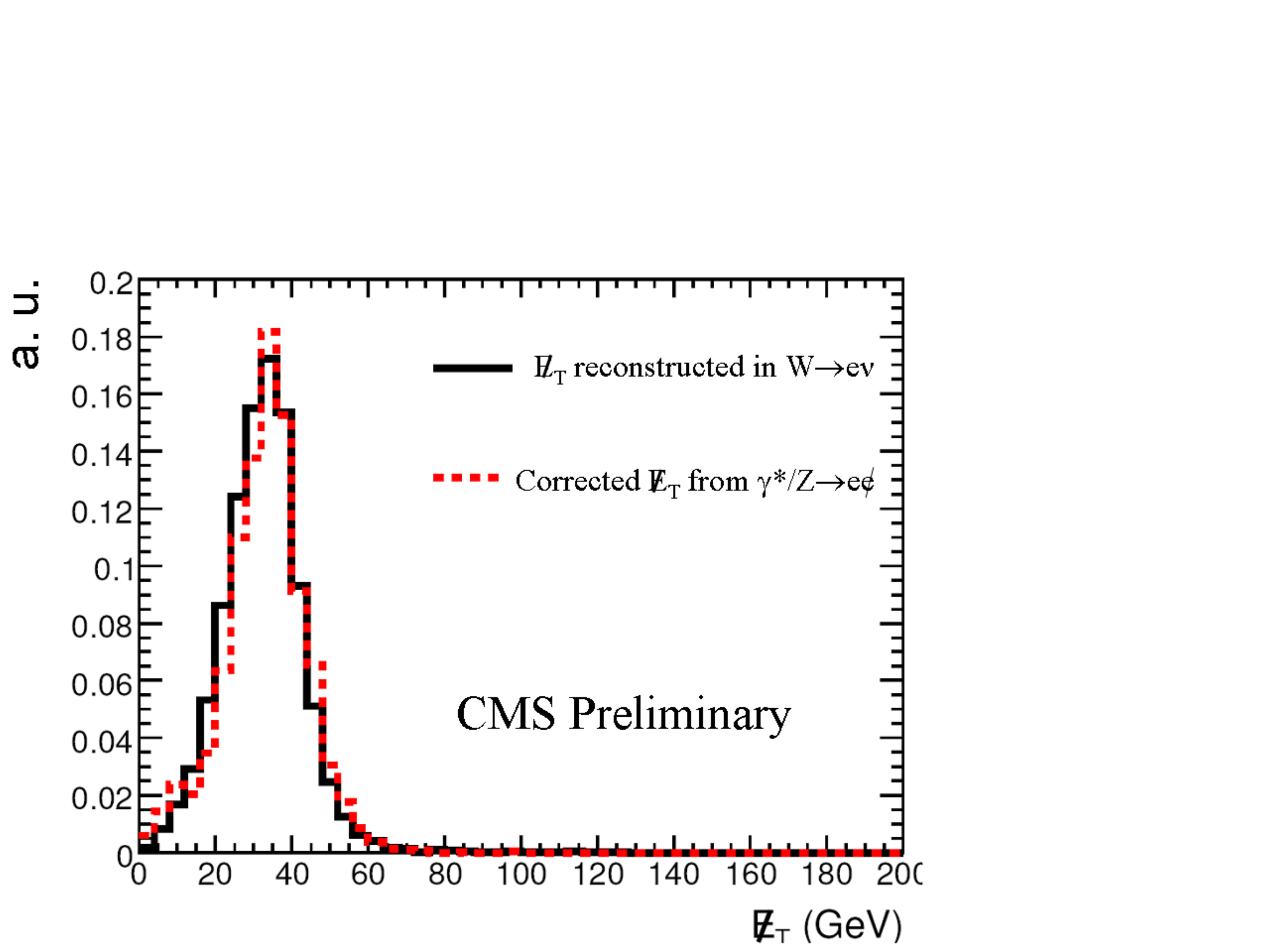}} &
   \resizebox{\linewidth}{!}{\includegraphics{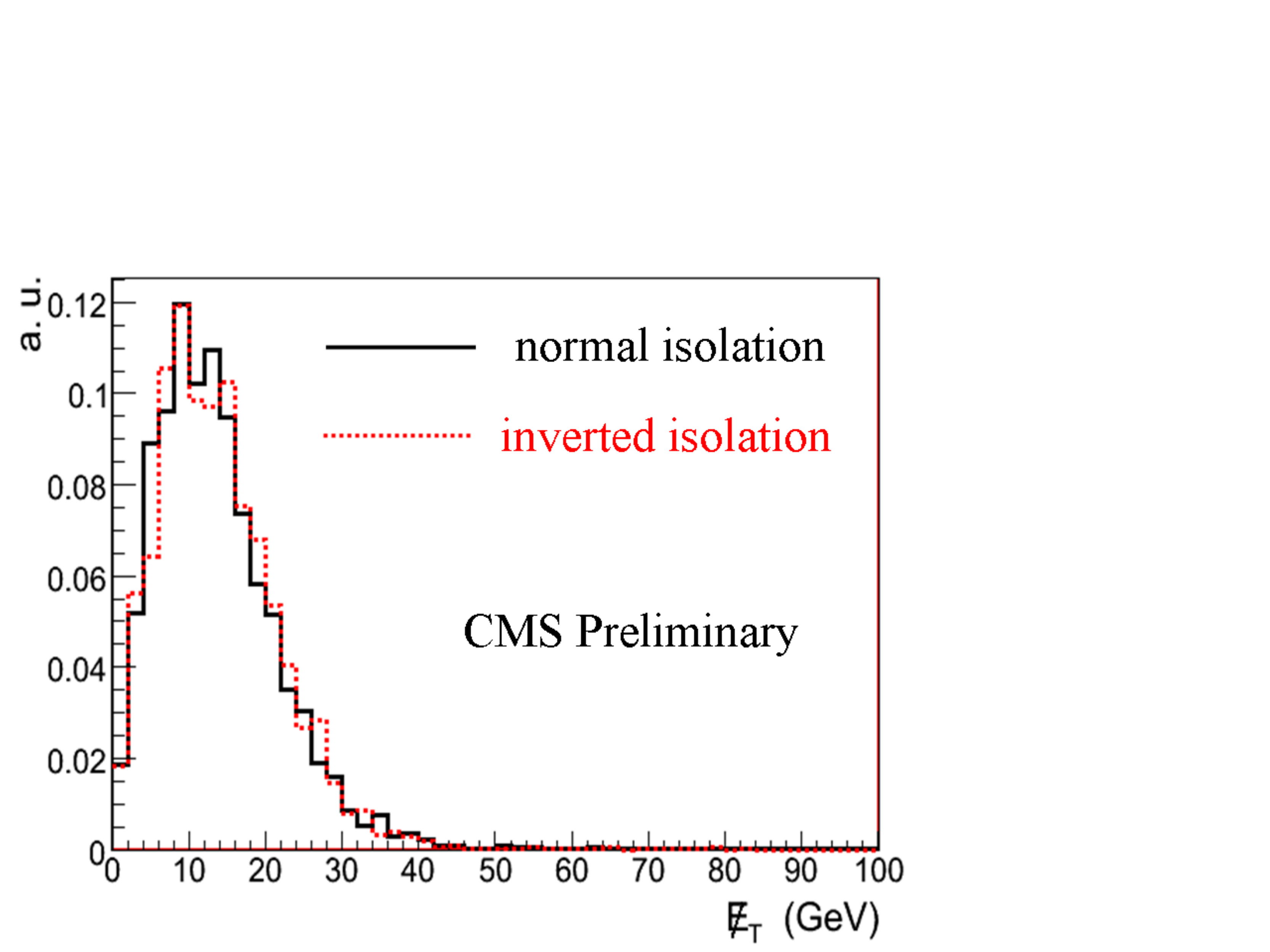}} \\
   \caption{The \met\ distribution of the selected $W\rightarrow e\nu$ sample (solid line) is well represented by the \Zee analogue (dashed line).}
   \label{f_Ersatz} &
   \caption{The \met\ distribution obtained by inverting one of the electron identification criteria represents that of the QCD di-jet background well.}
   \label{f_dijetTemplate} \\
 \end{2figures}
 \section{Cross-section Measurement}
The $\Wenu$ cross section is calculated using the following formula (similarly
for $\gammaZee$):
 \begin{equation}
  \sigma_{W}~\times~BR(W \rightarrow l\nu )~=~\frac{N_{W}^{sig}-N_{W}^{bkgd}}{A_{W}~\times~\epsilon_{W}~\times~\int Ldt}
 \label{eq:xsec}
 \end{equation}
 
 $N_{W}^{sig}$ and $N_{W}^{bkgd}$ are the number of signal and background events passing the selection. $\epsilon_{W}$ is the efficiency of the triggering, reconstruction and selection of the $W\rightarrow l\nu$ events. All are measured from data using the methods described. $A_{W}$ is the geometric and kinematic acceptance, which is determined from simulation. The integrated luminosity, $\int Ldt$, is measured externally to this analysis.

\section{Conclusions}
Analysis strategies for measuring the inclusive production cross-sections of the W and Z bosons have been formulated and tested for the early data-taking period of CMS. These strategies use robust selections and data-driven methods to extract efficiencies and background-corrected signal yields. This mitigates the effects of imprecise knowledge of the alignment and calibration of the detector and the impact of possibly inaccurate detector simulations. 

Significant results can be obtained with only 10\,pb$^{-1}$ of data.

\end{document}